\documentclass[showpacs,floatfix,eqsecnum,preprint,nofootinbib]{revtex4}
\usepackage{graphics}

\begin{document}

\title{Possible New Interactions of Neutrino and the KATRIN Experiment}
\author{A.~Yu.~Ignatiev}
\email{a.ignatiev@physics.unimelb.edu.au}
\author{B.~H.~J.~McKellar}
\email{b.mckellar@physics.unimelb.edu.au} \affiliation{
    ~{\em School of Physics, Research Centre for High Energy Physics,
     University of Melbourne,}\\
    {\em   Australia.}}

\def\be{\begin{equation}}
\def\ee{\end{equation}}
\def\bea{\begin{eqnarray}}
\def\eea{\end{eqnarray}}
\newcommand{\nn}{\nonumber \\}

\begin{abstract}
We analyse the possible role of new interactions of neutrino in the
forthcoming tritium  beta decay experiment KATRIN aimed at detecting the
neutrino mass with the sensitivity of 0.3 - 0.2 eV.

It is shown that under certain circumstances the standard procedure of data analysis would have to be modified by the introduction of an extra parameter describing the strength of the new interactions.

Our model simulations show that the modified procedure may improve the quality of the fit compared with the standard case. Ignoring the possibility
of new interactions may lead to a systematic error in the neutrino mass determination.
\\
\end{abstract}
\maketitle
\begin{section}{Introduction}

The compelling evidence for non-zero neutrino mass has been a recent 
triumph of modern science. Neutrino oscillation experiments give us 
information on squared mass differences between different types of 
neutrinos. However, the absolute values of neutrino masses remain 
unknown. One way to find the absolute mass is to study the electron 
spectrum in beta decay. As suggested by Fermi in the late thirties, the 
deviation of the linearised spectrum (the Curie plot) from a straight 
line near the end  point is a signal of non-zero neutrino mass. 

This idea has been implemented in a number of recent experiments 
\cite{ma,l} with all results consistent with zero mass and thus providing 
an upper limit on the neutrino mass. The work on the next generation   
tritium beta decay experiment, KATRIN, is in progress \cite{k,b}. 

Theoretically, the existence of neutrino mass and the existence of new neutrino interactions are closely related. This is because the neutrino interactions described by the Standard Model cannot generate the neutrino mass while the additional interactions can. Thus a question arises: what is the potential effect of new interactions in beta decay and how is the neutrino mass measurement is influenced?

This question has a long history starting from the time before V-A theory
was established. Obviously there was a need to analyse all possible types
of neutrino interactions in order to choose the one that was consistent
with experiment. More recently, the interest in this problem was revived
in \cite{sg,sgm,mgs} motivated, in particular, by an unexpected 
experimental finding that the best fit for the squared neutrino mass 
turned out to be negative. 

It was shown that the account of possible new neutrino interactions, such 
as right-handed (vector and scalar) currents can significantly affect the 
measured value of neutrino mass. In particular, the new interactions can 
drive negative the value of $m^2$ extracted from experiment whereas the 
physical value $m^2$     must be   positive. 

In this paper we extend the analysis of \cite{sg,sgm,mgs} by using the 
fact that, from the point of view of tritirm beta decay experiments, the 
neutrino spectrum can be considered degenerate. This leads to the 
appearance  in the electron spectrum of only one extra parameter 
describing the strength of the new interactions. The modified electron 
spectrum can be computed analytically. 

Thus, we are able to simulate the observed spectrum assuming that new interaction are present and have strength allowed by the existing constraints. We then can fit the simulated data by the usual spectral function (i.e. as if there were no new interactions). In such a procedure, the difference between the input and output values of the neutrino mass will                    describe the effect of new interactions on the neutrino mass measurement.
\end{section}
\begin{section}{Tritium beta decay spectrum in the presence of new interactions}
\begin{widetext}
As was shown in \cite{sgm}, the integral spectrum for tritium beta decay
near the end point is given by the following expression (the meaning of
the effective neutrino mass $m$ has recently been discussed in \cite{smi}):
\be\label{N}
 N(E)=K\left\{\frac{1}{3}(E^2-m^2)^{3/2}+xm
\left[E\sqrt{E^2-m^2}-m^2 \ln
\left(\frac{E+\sqrt{E^2-m^2}}{m}\right)\right] \right\}. \ee
 Here, $E$ is the maximum neutrino energy, $x$ is the dimensionless parameter describing the strength of
the new interactions (for consistency with previous works our notations
follow that of \cite{sgm,mgs}) :
\be
x=x_R+x_{SR}\ee
\be
x_R=-2\rho_R\frac{m_e}{\langle E
\rangle}\sum_i\cos\theta_i\cos\theta_{iR}\ee
\be
x_{SR}=-2\rho_{SR}\large(\frac{G_V^2}{G_V^2+3G_A^2}\large)
\sum_i\cos\theta_i\cos\theta_{iSR}\ee

Here, $x_R$ and $x_{SR}$ describe the strengths of right-handed and
scalar right-handed interactions, respectively; they  are expressed in
terms of other two convenient parameters with the same physical meaning,
$\rho_R$ and $\rho_{SR}$:
\be
\rho_R=\frac{g_R^2}{g^2}\frac{M_W^2}{M_R^2}(ME)_R \ee
\be
\rho_{SR}=\frac{g_{SR}^2}{g^2}\frac{M_W^2}{M_{SR}^2}(ME)_{SR},\ee where
the three sets $(g, M_W)$; $(g_R, M_R)$, and $(g_{SR}, M_{SR})$ refer to
the coupling constants/boson mass values for the standard, right-handed,
and the scalar right-handed interactions, correspondingly.

The factors $(ME)_R$ and $(ME)_{SR}$ account for the ratio of the
hadronic matrix elements of the  currents involved in tritium beta decay
relative to those of the Standard Model. Each factor includes the
elements of the quark CKM-type matrix generated by the appropriate
non-standard interaction.

Further, there are 3 sets of angles in the above formulas: $\theta_i$,
$\theta_{iR}$, and $\theta_{iSR}$ where index $i$ running over  1,2,3
refers to the neutrino mass eigenstates. The angles $\theta_i$ belong to
the Standard Model and arise because the standard weak interaction
eigenstates are different from mass eigenstates. Similarly, if a new
interaction is introduced, a new set of angles arises for the same
reason. Although in general these new sets would be different from the
Standard Model set (and from each other) it is usually assumed for
simplicity that they are the same, i.e.
\be
 \theta_{iR}=\theta_{iSR}=\theta_i.\ee
Under this assumption the existing experimental constraints \cite{x} on 
$\rho_R$ and $\rho_{SR}$ are \footnote{In deriving these constraints  the 
general approach (see e.g. \cite{lr}) is followed and it is not assumed 
that the right-handed quark mixing angles are equal to the left-handed 
ones.} 
\be
\rho_R\leq 0.07\;\;\;\; \rho_{SR}\leq 0.1.\ee

Finally, plugging these into formulas for $x$ (and assuming that
$m_e/\langle E\rangle \sim 1)$ we obtain:
\be
|x_R| \leq 0.14\;\;\;\; |x_{SR}|\leq 0.035. \ee
\end{widetext}
\end{section}
\begin{section}{Model simulations}

In the context of the KATRIN experiment we have conducted a study of possible
role of the new interactions by carrying out a  number of simulations
using Mathematica \cite{m} as our tool.

To generate out simulated data we used the theoretical formula  (\ref{N}) 
plus the  ``random error'' term normally distributed around zero with the 
dispersion $s$. The value of $s$ was determined by a self-consistency 
requirement imposed by the KATRIN conditions: at the input value of 
neutrino mass $m= 0.35$  $eV$ the $1\sigma$ statistical error in the mass 
determination should be 0.07 $eV^2$ \cite{b}. Typically, we used as input 
values $x=\pm{0.14}$ and  $m= 0.35$  $eV$. 

Regarding the energy, we assumed that E can take 20 values starting from
E = 1 eV through to E = 20 eV with a step of 1 eV. We have tried several
methods of extracting the neutrino mass from our simulated data.

In Method A  we generated data according to Eq.(\ref{N}) with non-zero 
$x$ and $m= 0.35$  $eV$, and then did an analysis assuming $x=0$ and 
finding the best fit for $m$ or $m^2$ based on Eq.(\ref{N}) with $x=0$. 

In Method B the data were generated in the same way as in Method A, but 
as our fitting formula we used Eq.~(\ref{N}) with 2 parameters ($m$ and 
$x$) to be fitted. 

In Method C the data generation method was again the same as above, but 
for the fitting purposes the parameters $m$ and $x$ were treated 
``asymmetrically'':  the neutrino mass was considered as a fitting 
parameter while $x$ took on different but fixed values. 

Method A for $x=-0.14$ yields $m=1 $ $eV$ (with negligible dispersion), 
i.e. it leads to a large (about 0.7 eV) systematic error in neutrino mass 
determination. The very small dispersion  in $m$ values is related to the 
fact that in our procedure values of $m$ larger than  $1 $ $eV$ are not 
allowed because for $E=1$ $eV$ they   lead to negative values under the 
square roots in Eq.~(\ref{N}). Thus values we can investigate are 
squashed against this limit.  In addition, the quality of the fit turns 
out to be bad ($\chi^2$/d.o.f. $\simeq 6 $). For $x=+0.14$ the method 
yields negative $m^2$ values: $m^2=-1.39\pm 0.08$, with $\chi^2$/d.o.f. 
$\simeq 2.7 $. 

Method B was used only for  $x=-0.14$.  For $x=+0.14$, based on results 
of Method A,  one would expect that the best fit can yield negative $m^2$ 
values, but the fitting formula, Eq.~(\ref{N}), contains not only $m^2$, 
but also $m$. Therefore it cannot be used in the case of  negative 
$m^2$ without modifications.   

Method B gives much better $\chi^2$/d.o.f. values than method A 
($\chi^2$/d.o.f. $\simeq 0.9 $) . We ran 10 simulations, with the results 
plotted in figure 1. These results give  the average output values 
$m=0.65 \pm 0.11$ $eV$ and $x=-0.06\pm 0.017$.  It is disappointing that the mean 
values for our simulation are more than 3 standard deviations away from 
the input values.  This can be related to the fact that the contours of 
equal $\chi^2$ in the plane ($x$, $m$) do not enclose a small region
 as can be seen in 
Fig.~\ref{1}. Indeed, there seems to be a very long valley in the vicinity of the minimum.
\begin{figure}
\includegraphics{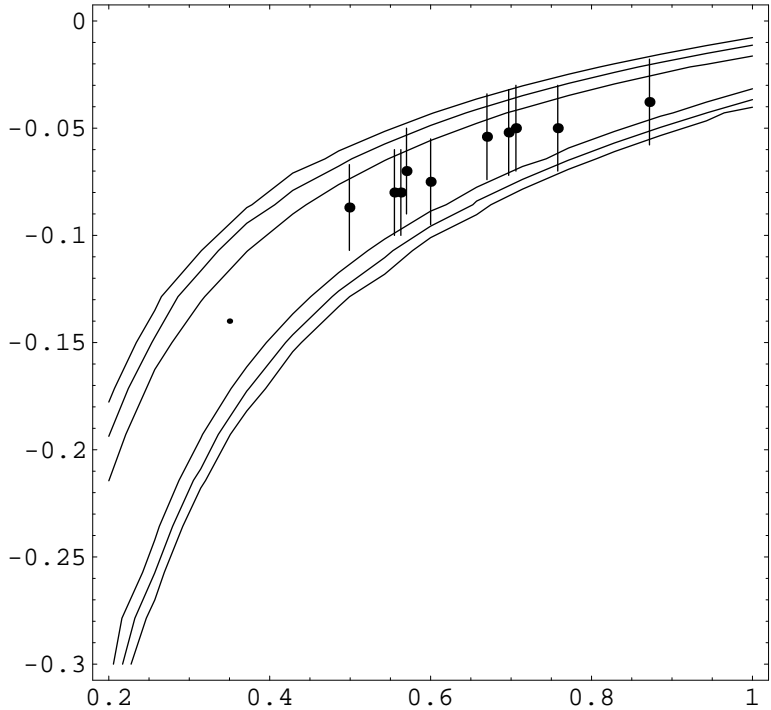}
\caption{\label{1}Contours of equal $\chi^2$ in the plane ($m$, 
horizontal axis, vs. $x$, vertical axis). The $\chi^2 $ values on the contours, moving out of the minimum valley, are $\chi^2/dof=1.9, 2.9, 3.9$. The value of $m$ ranges between 
0.2 and 0.4. The dot without error bars is the input value.}   
\end{figure}

  Finally, Method 
C (see Table 1) seems to better reproduce the value of $m$; its 
unpleasant feature is that  without additional information we do not know 
in advance the value of the $x$ 
parameter that should be plugged in before the fitting starts. Because of 
that, one can start with the largest (by modulus) value of $x$ allowed by 
the modern data and then gradually lower this value and see if the 
quality of the fit ($\chi^2$) has improved. However, from Table 1 we see 
that a good fit can be obtained for a range of $x$ values, and the 
outstanding problem is how to narrow this range down. The challenge will 
be to obtain independent measurements of, or limits on, $x$. Note also, 
that for the same reason as in Method B, only negative values of $x$ were 
used. 

Therefore, if the standard procedure of KATRIN data analysis produces a
fit that is not good enough and/or the best fit for the neutrino mass
turns out to be unphysical then the hypothesis of new interactions should
be tested as described above.
\begin{table}
\caption{\label{t}Method C with input values $x=-0.14$, $m=0.35$ $eV$.} 
\begin{tabular}{|c|c|c|}\hline Trial value of $x$ & Output value of $m$, $eV$ 
& $\chi^2$/d.o.f.
\\ \hline 0 & $1$ & 6 
\\
 -0.10 & $0.46 \pm 0.02$ & 0.9 \\ 
-0.15  & $0.33 \pm 0.01$ & 0.9
\\ -0.20 & $0.25 \pm 0.01$ & 0.9\\  \hline 
\end{tabular}
\end{table} 
We have performed additional simulations with the same input values of $m$ and $x$,
but with smaller dispersions $s$. We find that, with a dispersion corresponding to a $1\sigma$
statistical error in the mass determination of 0.015 $eV^2$, reliable values of $x$ and $m$ can be extracted.
\end{section}
\begin{section}{Conclusions}
We have analysed the possible role of new interactions of neutrinos in the
forthcoming tritium  beta decay experiment KATRIN aimed at detecting the
neutrino mass with the sensitivity of 0.3 - 0.2 eV.

It is shown that under certain circumstances the standard procedure of
data analysis would have to be modified by the introduction of an extra
parameter describing the strength of the new interactions.

Our model simulations show that the modified procedure may improve the
quality of the fit compared with the standard case. We find that it is possible for the new interactions, if present, to lead 
to a systematic error in the mass determination from an analysis which ignores this presence.
However when new interactions are included in the analysis the mass determination may still be unreliable unless

(i) the strength of the new interaction can be determined by independent experiments and used as an 
input parameter in the analysis of the experiment, or

(ii) the statistical (and systematic) errors in the experiment can be reduced
  to 0.015 $eV^2$ in the mass at 1 $\sigma$.

We recognise that our simulations do not include all of the details necessary for a full 
simulation of these effects in the KATRIN experiment. But our results indicate that such a 
simulation could usefully be undertaken.
\end{section}
\begin{section}{Acknowledgements}
This work was supported in part by the Australian Research Council.
\end{section}

\end{document}